\documentclass[prl, twocolumn, superscriptaddress,showpacs]{revtex4}

\usepackage{bm}
\usepackage{epsfig}

\begin{document}

\title{Hidden topological structure in the continuous Heisenberg spin chain }

\author {R. Dandoloff}
\email{rossen.dandoloff@ptm.u-cergy.fr}

\affiliation{ Laboratoire de Physique Th\'{e}orique et
Mod\'{e}lisation, Universit\'{e} de Cergy-Pontoise, F-95302
Cergy-Pontoise, France}

\begin{abstract}
In order to study the spin configurations of the classical
one-dimensional Heisenberg model, we map the normalized unit
vector, representing the spin, to a space curve. We show that the
total chirality of the configuration is a conserved quantity. When
the space curve forms a knot, this defines a new class of
topological spin configurations for the Heisenberg model.

\end{abstract}


\maketitle It is well known that the two-dimensional continuous
Heisenberg model has very nice topological properties (Belavin and
Polyakov 1975) \cite{BP}. The order parameter is a normalized
vector field ${\bf n}^2=1$ (therefore the order parameter manifold
is $S^2$). If we impose homogenous boundary conditions on the
vector field ${\bf n(r)}_{r\to\infty}={\bf n}_0$ ( constant vector
field ), we can compactify the plane $R^2$ into $S^2$ and
therefore the possible field configurations are classified by
$\pi_2(S^2)=\bf Z$. The energy in each class is bounded from below
$H\ge nJ$, where $n$ is the number of times $S^2$ is wrapped
around $S^2$, and $J$ is the coupling constant in the Heisenberg
spin hamiltonian. Unfortunately the one dimensional Heisenberg
model does not have this nice topological property. Under
homogeneous boundary conditions the line $R^1$ may be compactified
to $S^1$, and now $\pi_1(S^2)=0$ and there are no different
classes of configurations based on homotopy. In order to find out
if there is a hidden topological structure in the one dimensional
case one has to analyze the Heisenberg hamiltonian in more
details. The vector field is normalized and therefore we will use
the following representation for ${\bf
n}=(\sin\theta\cos\phi,\sin\theta\sin\phi,\cos\theta)$. In
$\theta$ and $\phi$ variables the hamiltonian has the form:
$$ H=J\int_{-\infty}^{+\infty}(\theta_s^2 +
\sin\theta^2\phi_s^2)ds \eqno(1)$$

where the subscript $s$ stands for $\frac{d}{ds}$ and $s$ denotes
the coordinate along $R^1$. This hamiltonian is not symmetric
under homotety transformation $s\to\lambda s$ and therefore the
spin configurations are not metastable like in the $2D$ case. The
equations of motion for this spin hamiltonian have been
established (Tjon and Wright 1977)) \cite{TW} in taking $\phi$ and
$\cos\theta$ to be the conjugated generalized coordinate and
momentum so that the Poisson bracket gives
$[\phi(x),\cos\theta(y)]=\delta(x-y)$. The generator of
translations (momentum) is given by the following expression (Tjon
and Wright 1977) \cite{TW}:
$$ P=\int_{-\infty}^{+\infty}(1-\cos\theta)\phi_s ds \eqno(2)$$

where the third component of the normalized generator of rotations
(magnetization) is given by (Tjon and Wright 1977)\cite{TW}:
$$ M=\int_{-\infty}^{+\infty}(\cos\theta - 1)ds \eqno(3) $$

The quantities $P$ and $M$ are constants of the motion. For our
analysis of the possible spin configurations it is useful to map
the unit vector $\bf n$ to the unit tangent of a space curve
(Balakrishnan et al. 1990) \cite{BBD}. Now different space curves
will represent different spin configurations. We will impose
homogeneous boundary conditions, which will assure that the energy
is finite and the curves representing the different spin
configurations will tend to the straight line as $s\to \pm\infty$
Now we will concentrate on the geometrical and topological
quantities characterizing a space curve. Of special interest for
us will be the writhe of a curve (which characterizes the
chirality of the curve). It is defined as follows:
$$ Wr=\frac{1}{4\pi}\int_{-\infty}^{+\infty} ds\int_{-\infty}^{+\infty}
 ds' \frac{(\bf r - \bf r').(\bf n
-\bf n')}{|\bf r - \bf r'|} \eqno(4) $$

The tip of the radius vector $\bf r$ draws the curve, while $\bf
n$ is the unit tangent. A theorem by Fuller (Fuller
1978)\cite{Ful} allows to express $Wr$ as an integral of a local
quantity. We will express $Wr$ with respect of a reference curve
$C_0$( Fain and Rudnick 1997)\cite{DNA}:
$$ Wr=Wr_0 + \frac{1}{2\pi}\int_{-\infty}^{+\infty}\frac{{\bf n}_0\times{\bf
n}.\frac{d}{ds}({\bf n}_0+{\bf n})}{(1+{\bf n}_0.{\bf n})}ds
\eqno(5) $$
where $Wr_0$ is the writhe of the reference curve. The
simplest choice is the straight line $C_0=(0,0,s)$, then ${\bf
n}_0=(0,0,1)$ and $Wr_0=0$( Fain and Rudnick 1997)\cite{DNA}. A
simple calculation gives the following expression for the writhe:
$$ Wr=\frac{1}{2\pi}\int_{-\infty}^{+\infty}(1-\cos\theta)\phi_s ds \eqno(6)$$

Our first observation is that the writhe $Wr$ for the spin
configurations (quantity that characterizes the chirality of the
spin configuration) coincides with the total momentum $P$. The
total momentum $P$ is a conserved quantity - it follows that $Wr$
is a conserved quantity too. This will lead us to a new class of
possible excitations for the continuous classical spin Heisenberg
model. We will note first that the writhe $Wr$ suffers
discontinuity when one region of the curve crosses another and the
jump is always +2(Frank-Kamenetskii and Vologodskii)\cite{Uspehi}.
This means that all configurations that belong to the
configuration of the ground state ($\theta=0$) are separated from
all other classes of configurations by a jump of the writhe $Wr$.
Let us consider one such configuration: the space curve
representing the spin configuration forms a knot with a loop which
follows the semi-circle at infinity and then comes back from
$s=\pm\infty$ as a straight line and goes into the actual knot.
One can imagine also a knot which is cut at $s=s_0$ and then both
ends are pulled to $+\infty$ and $-\infty$ and are put together
over the infinite semi-circle (see Fig.1). The writhe is zero for
both straight segments when $s\to\pm\infty$ and for the infinite
semi-circle. This geometrical construction does not change the
writhe of the actual knot. Such a knot belongs to a whole class of
configurations which deform smoothly from one to another and who
are separated from the ground state class by a jump in the writhe
$Wr$. Belonging to the knot configuration will have consequences
for the energy of the spin configuration too. Let us consider the
following Cauchy-Schwarz inequality:
$$
\int_{-\infty}^{+\infty}\sin^2\frac{\theta}{2}
ds\int_{-\infty}^{+\infty}\phi_s^2\sin^2\frac{\theta}{2} \geq $$
$$\geq\left(\int_{-\infty}^{+\infty}\phi_s\sin^2\frac{\theta}{2} ds
\right)^2=\frac{P^2}{4} \eqno(7)
$$

The energy satisfies the obvious inequality:
$$ J\int_{-\infty}^{+\infty}(\theta_s^2 + \sin^2\theta\phi_s^2)ds=J\int_{-\infty}^{+\infty}(\theta_s^2 +
4\sin^2\frac{\theta}{2}\phi_s^2)ds \geq$$
$$\geq J\int_{-\infty}^{+\infty}
4\sin^2\frac{\theta}{2}\phi_s^2 ds \eqno(8) $$

Combining inequalities (7) and (8) leads to the following
inequality for the energy:
$$ H\geq J\frac{P^2}{M} \eqno(9) $$

Let us note here that $M=0$ only for the ground state $\theta=0$
and that $P\neq 0$ for curves in the knot configuration. Thus the
energy is limited from below for such a configuration.

We have shown that there are topological configurations for the
Heisenberg spin model even in the one-dimensional case. One should
investigate the different knot configurations in order to
elaborate a classification of such configurations according to the
type of knot they represent.

\begin{figure}
  \includegraphics[width=2.5in]{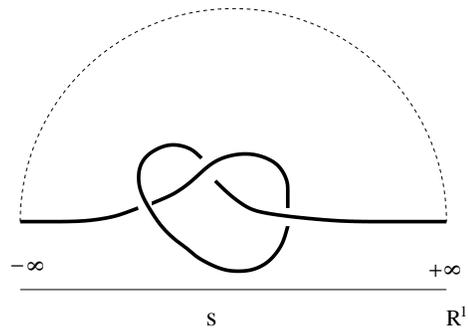}
  \caption{ Spin configuration}
\end{figure}

\end{document}